\documentclass[aps,prl,twocolumn]{revtex4-2}
\usepackage{amsmath,amssymb,bm}
\begin{document}

\title{Continuum Free-Energy Computing}

\author{Trey Li}
\affiliation{University of Manchester, Manchester, UK}
\date{March 28, 2026}

\begin{abstract}

Building on nonintrinsic Landau theory, we introduce continuum free-energy computing as a new computing paradigm in which problem instances are encoded in programmable free-energy functionals and solved by intrinsic relaxational dynamics. We identify ion-patterned FeRh as a plausible physical realization through spatial control of the local phase bias, with antiferromagnetic--ferromagnetic interface motion providing the relaxational mechanism. We further identify two representative task classes, a minimal operating protocol, and the main physical constraints.

\end{abstract}

\maketitle

\emph{Introduction.---}
Computation is traditionally formulated in a finite-dimensional state space.
Across digital computing \cite{HennessyPatterson}, quantum computing \cite{NielsenChuang}, analog computing \cite{MacLennan2009}, neuromorphic computing \cite{Indiveri2011}, reservoir computing \cite{Lukosevicius2009}, quantum annealing \cite{Kadowaki1998}, and Ising-based optimization \cite{Yamamoto2017}, problem instances are encoded in a finite set of degrees of freedom, and computation proceeds through dynamics on those degrees of freedom.

At the same time, condensed matter physics has long studied another class of dynamics that evolves in an infinite-dimensional space. Many systems relax as continuous fields governed by free-energy functionals \cite{CrossHohenberg1993,Chen2002}. Standard examples include phase ordering, pattern formation, and domain evolution. These dynamics are usually studied as physical phenomena rather than as computation.

In \cite{li2026}, we showed that externally written microscale fields can survive coarse graining and appear in the coarse-grained free-energy functional as spatially prescribed coefficients. This defines a nonintrinsic sector of Landau theory, in which part of the functional is no longer purely intrinsic but externally writable.

That observation raises a natural question. Can problem instances be encoded directly in a free-energy functional, so that relaxational dynamics performs the computation? Here we show that, in principle, this is possible. We call the resulting paradigm \emph{continuum free-energy computing} (CFEC).

CFEC differs from Ising machines and quantum annealers, which encode discrete Hamiltonians over binary variables \cite{Kadowaki1998,Yamamoto2017}, from neuromorphic and reservoir platforms, which evolve continuous signals on finite networks \cite{Indiveri2011,Lukosevicius2009}, and from conventional phase-field systems, which minimize free energies without usually treating the coefficients of the functional itself as the programmable carrier of problem instances \cite{CrossHohenberg1993,Chen2002}. In CFEC, computation takes place directly in a programmable free-energy functional over a continuum field.

\emph{Computational framework.---}
CFEC has three elements: encoding, relaxation, and readout. A problem instance $P$ is first encoded in a spatially programmable field $f_{P}(\mathbf r)$, which defines a free-energy functional $F_{P}[m]$ of a continuous order parameter $m(\mathbf r)$. The dominant mesoscopic dynamics of the material then lower $F_{P}$. Finally, a readout map converts the relaxed configuration $m_\ast(\mathbf r)$, or the domain geometry it defines, into a solution object such as a binary assignment or an interface geometry.

A physical realization of CFEC therefore requires three conditions. First, the problem instance must enter through externally writable coefficient fields that survive coarse graining and are not erased by internal reconstruction. Second, the dominant dynamics must be able to lower the programmed free energy without an externally imposed update sequence during operation. Third, the relaxed configuration must be experimentally readable and sufficiently stable for reliable readout.

These three conditions have natural counterparts in ion-patterned FeRh. The first is provided by irradiation-written coefficient fields that generate spatially programmable phase bias, as shown in \cite{li2026} and made experimentally plausible by \cite{Bali2014,Heidarian2015,Cress2021}. The second is provided by antiferromagnetic--ferromagnetic (AF--FM) relaxational dynamics in the coexistence regime \cite{LeGraet2015}, which lower the programmed free energy. The third is provided by magnetic-domain contrast, for example through XMCD-PEEM imaging \cite{Baldasseroni2012}, which makes the relaxed configuration experimentally readable.
We therefore illustrate CFEC in the concrete setting of ion-irradiated FeRh.

\emph{Programmable functional.---}
To formulate CFEC in FeRh, we use a coarse-grained scalar field $m(\mathbf r)$ to distinguish mesoscopic AF-like and FM-like regions and their interfaces. Here $m(\mathbf r)$ should be understood as an effective phase-indicator field rather than as a microscopic magnetic order parameter. It parametrizes the local mesoscopic phase state, while its spatial configuration constitutes the computational variable.

As shown in \cite{li2026}, once a written microscale pattern survives coarse graining and remains inert on the timescale of interest, it enters the coarse-grained free-energy functional as a spatially varying coefficient field. We therefore write the local free-energy density in the form
\begin{equation*}
f_{\mathrm{loc}}(m,\mathbf r)=V(m)+a(\mathbf r)u(m), 
\end{equation*}
where $V(m)$ is the intrinsic local potential, $a(\mathbf r)$ is an externally written coefficient field, and $u(m)$ specifies how that field couples to the local phase state. A simple choice is
\begin{equation*}
V(m)=\frac{\lambda}{4}(m^2-1)^2,
\qquad
u(m)=m,
\end{equation*}
with $\lambda>0$. Then $V(m)$ provides two local minima corresponding to the competing phase states, while the written field $a(\mathbf r)$ shifts their relative stability in space through the term $a(\mathbf r)m$.

In FeRh, we control $a(\mathbf r)$ by controlling the local AF--FM transition temperature $T_{\mathrm t}(\mathbf r)$ through ion irradiation \cite{Bali2014,Heidarian2015,Cress2021}. 
The resulting phase behavior can be spatially structured, and depth-selective phase coexistence has also been observed in FeRh thin films \cite{Griggs2020}. At a fixed operating temperature $T$, the sign of $T-T_{\mathrm t}(\mathbf r)$ indicates which phase is locally favored: regions with $T>T_{\mathrm t}(\mathbf r)$ favor FM-like order, whereas regions with $T<T_{\mathrm t}(\mathbf r)$ favor AF-like order. We therefore parametrize the written coefficient field as
\begin{equation*}
a(\mathbf r)=\alpha\bigl[T-T_{\mathrm t}(\mathbf r)\bigr],
\end{equation*}
where $\alpha$ is a phenomenological constant. Programming $a(\mathbf r)$ therefore reduces to writing $T_{\mathrm t}(\mathbf r)$.

To obtain the full mesoscopic functional, one must also include the energetic cost of interfaces and, when relevant, additional weak long-range couplings. Interfacial effects are captured by
\begin{equation*}
F_{\mathrm{grad}}
=
\frac{\kappa}{2}\int |\nabla m|^2\,d\mathbf r,
\end{equation*}
which penalizes sharp spatial variations and sets a finite interface width. Additional weak long-range interactions may be included through
\begin{equation*}
F_{\mathrm{nl}}
=
\frac12\iint K(\mathbf r-\mathbf r')\,m(\mathbf r)m(\mathbf r')\,d\mathbf r\,d\mathbf r',
\end{equation*}
which can represent weak magnetostatic or elastic couplings. The total programmable functional is therefore
\begin{equation*}
F[m]
=
\int f_{\mathrm{loc}}(m,\mathbf r)\,d\mathbf r
+F_{\mathrm{grad}}+F_{\mathrm{nl}}.
\end{equation*}
Problem instances are encoded through the written coefficient field $a(\mathbf r)$.

\emph{Relaxation and sharp-interface limit.---}
When the temperature is set in the AF--FM coexistence window, the system evolves through nucleation and interface motion. At mesoscopic scales, this slow evolution is well approximated by Model-A dynamics \cite{HohenbergHalperin1977},
\begin{equation*}
\partial_t m(\mathbf r,t)
=
-\Gamma\frac{\delta F}{\delta m(\mathbf r,t)}
+\eta(\mathbf r,t),
\end{equation*}
where $\Gamma$ is the relaxation rate and $\eta$ is thermal noise. Once a problem instance has been encoded in $F[m]$, the subsequent physical relaxation becomes the computation.

In the noiseless limit $\eta=0$,
\begin{equation*}
\frac{dF}{dt}
=
-\Gamma\int
\left|\frac{\delta F}{\delta m}\right|^2 d\mathbf r
\le 0,
\end{equation*}
so the programmed free energy decreases monotonically with time. The system therefore moves downhill on the written free-energy landscape until it reaches a stable or metastable configuration from which local relaxational motion can no longer lower $F$.

In the coexistence regime, the same dynamics often admit a sharper description in terms of AF--FM domain walls. When the typical domain size is much larger than the interface width, the free energy is approximated by a bulk term plus an interface term \cite{Chen2002},
\begin{equation*}
F
\approx
\sum_i \int_{\Omega_i} f_{\mathrm{loc}}\,d\mathbf r
+\sigma \mathcal L_{\mathrm{DW}},
\end{equation*}
where $\{\Omega_i\}$ are the domains, $\mathcal L_{\mathrm{DW}}$ is the total domain-wall length, and $\sigma$ is the interfacial energy per unit length. The relaxed pattern is therefore determined by a competition between local phase preference and the tendency to keep the total interface length small.

When the interface width is much smaller than both the interface curvature radius and the spacing between neighboring interfaces, the interface motion is well captured schematically by the local normal-velocity law \cite{Allen1979,Chen2002}
\begin{equation*}
v_n
=
-\Gamma\Bigl(
\sigma \kappa_{\mathrm{geom}}
+\Delta f_{\mathrm{loc}}
+H_{\mathrm{nl}}
\Bigr),
\end{equation*}
where $\kappa_{\mathrm{geom}}$ is the interface curvature, $\Delta f_{\mathrm{loc}}$ is the local free-energy difference across the interface, and $H_{\mathrm{nl}}$ collects weak nonlocal contributions. Curvature tends to shorten and smooth the interface, while the written coefficient field enters through $\Delta f_{\mathrm{loc}}$ and provides the programmed driving force. In the coexistence regime, computation therefore takes the form of domain-pattern relaxation on a written free-energy landscape.

\emph{Readout.---}
The relaxational dynamics produce a final physical configuration rather than a mathematical solution by themselves. Readout is the step that extracts the output from that configuration. For FeRh, the most direct readout channel is magnetic contrast. In particular, magnetic imaging methods such as XMCD-PEEM have been used to observe the nucleation, growth, and coexistence of magnetic domains across the AF--FM transition in FeRh thin films \cite{Baldasseroni2012,Temple2018,Almeida2017}. This makes the relaxed configuration experimentally accessible as a spatial object.

\emph{Representative task classes.---}
We now illustrate how problem instances are encoded into a written transition-temperature pattern and solved by relaxational dynamics in CFEC. The examples below concern tasks directly expressible through the local writing channel $a(\mathbf r)=\alpha[T-T_{\mathrm t}(\mathbf r)]$.

\textit{Example 1: A minimal discrete problem.---}
Given real numbers $h_1,\dots,h_n$, the problem is to find a binary vector $s=(s_1,\dots,s_n)\in\{-1,+1\}^n$ minimizing $\sum_{i=1}^n h_i s_i$.

To represent this problem in CFEC, partition the film into pairwise disjoint cells
\begin{equation*}
\Omega=\bigcup_{i=1}^n \Omega_i \qquad
\Omega_i\cap\Omega_j=\varnothing \text{ for } i\neq j.
\end{equation*}
For each $i\in\{1,\dots,n\}$ and each $\mathbf r\in\Omega_i$, set
\begin{equation*}
T_{\mathrm t}(\mathbf r)=T_{\mathrm t}^0+\Delta T_i,
\qquad
\Delta T_i=\beta h_i,
\end{equation*}
where $\beta\in\mathbb R$ is an encoding scale factor fixed by calibration. It must be large enough that different inputs produce distinguishable local biases, yet small enough that the corresponding temperature shifts remain inside the writable experimental range. Then for $\mathbf r\in\Omega_i$ one has
\begin{equation*}
a(\mathbf r)=a_i
=
\alpha\bigl[T-T_{\mathrm t}^0-\beta h_i\bigr]. 
\end{equation*}
After writing $T_{\mathrm t}(\mathbf r)$, bring the sample into the coexistence regime and let it relax. Define the cell averages
\begin{equation*}
\bar m_i=\frac{1}{|\Omega_i|}\int_{\Omega_i}m(\mathbf r)\,d\mathbf r
\end{equation*}
and read out
\begin{equation*}
s_i=\operatorname{sign}(\bar m_i)\in\{-1,+1\}.
\end{equation*}
Thus $s_i=+1$ if cell $\Omega_i$ relaxes predominantly to the FM-like phase and $s_i=-1$ if it relaxes predominantly to the AF-like phase.

\textit{Example 2: A continuum separation problem.---}
Given a bounded domain $\Omega\subset\mathbb R^2$ and disjoint prescribed subsets $A,B\subset\Omega$, the problem is to find a separating interface or a measurable set $D\subset\Omega$ such that $A\subset D$, $B\subset\Omega\setminus D$, and
\begin{equation}
F_{\mathrm{sep}}[D]
=
\int_D w(\mathbf r)\,d\mathbf r
+
\sigma\,\mathrm{Per}(D)
\label{eq:Fsep_task}
\end{equation}
is minimized over all such admissible sets $D$. Here $w(\mathbf r)$ is a prescribed spatial weight field, $\mathrm{Per}(D)$ is the perimeter of $D$, and $\sigma>0$ is the interfacial cost per unit length.

In CFEC, the unknown set is represented by the FM-like region
\[
D=\{\mathbf r\in\Omega:m(\mathbf r)>0\}.
\]
Choose a local bias field that favors the FM-like phase on $A$ and the AF-like phase on $B$, for example
\begin{equation*}
w(\mathbf r)=-w_0 \ \text{on } A,
\qquad
w(\mathbf r)=+w_0 \ \text{on } B,
\end{equation*}
with $w_0>0$, and weak or neutral bias elsewhere. Encode this by
\begin{equation*}
a(\mathbf r)=\gamma\, w(\mathbf r),
\end{equation*}
where $\gamma>0$ is a fixed calibration factor, or equivalently by
\begin{equation*}
T_{\mathrm t}(\mathbf r)=T-\alpha^{-1}\gamma\, w(\mathbf r).
\end{equation*}
After writing $T_{\mathrm t}(\mathbf r)$, bring the sample into the coexistence regime and let it relax. In the sharp-interface regime, the effective objective is Equation~(\ref{eq:Fsep_task}), and the relaxed domain $D_\ast$, equivalently its boundary $\partial D_\ast$, is the computed solution. 

\emph{Minimal protocol and physical constraints.---}
The FeRh construction discussed here is not a claim of a completed computing device. Rather, it defines a minimal physical architecture and experimental route for realizing the local-writing regime of CFEC. First, a spatial map of transition temperature $T_{\mathrm t}(\mathbf r)$ is written by ion irradiation, thereby defining the programmed coefficient field $a(\mathbf r)=\alpha[T-T_{\mathrm t}(\mathbf r)]$. Second, the sample is initialized in a uniform phase outside the coexistence window. Third, the operating temperature is brought into the AF--FM coexistence regime, where interfaces become mobile and the system evolves autonomously under relaxational dynamics. Finally, the resulting domain pattern is imaged and interpreted as the output associated with the encoded instance.

Because the effective functional is generally nonconvex, the dynamics do not in general guarantee the global minimizer. The output is instead a stable or metastable configuration selected by the competition among programmed bias, interfacial regularization, disorder, and thermal activation. CFEC should therefore be understood first as a physical variational primitive rather than as a guaranteed global optimizer.

A primary requirement for reliable operation is that the programmed energetic scale dominate uncontrolled disorder. Defects, pinning centers, thickness variations, and unintended compositional gradients all perturb the effective free-energy landscape and can exert quenched forces on interfaces. Thermal noise then plays a dual role. It can assist escape from shallow metastable states, but it also increases run-to-run variability. The operating temperature therefore sets a trade-off between exploration and reproducibility.

Several physical constraints delimit the accessible regime. The smallest programmable feature size is set by the irradiation resolution together with the requirement that imposed structure exceed the intrinsic correlation length \cite{li2026,Bali2014,Gatel2017,Almeida2020}. The readout resolution must in turn be sufficient to resolve the relaxed pattern on the scale at which the task is defined. Relaxation times are controlled by the mobility $\Gamma$ and the characteristic driving forces. In the sharp-interface regime, a typical interface velocity scales as
\[
v\sim \Gamma(\sigma/L+\Delta f_{\mathrm{loc}}),
\]
where $L$ is a typical domain scale.

A useful design guideline is the hierarchy
\[
\Delta F_{\mathrm{programmed}}
\gg
\Delta F_{\mathrm{disorder}}
\gtrsim
k_{\mathrm B}T,
\]
so that the written bias dominates uncontrolled disorder while thermal activation remains strong enough to assist escape from shallow traps. Within this operating window, ion-patterned FeRh provides a concrete minimal route toward CFEC. 

\emph{Conclusion.---} 
We have introduced continuum free-energy computing in which writable free-energy coefficients give relaxational condensed-matter dynamics a computational interpretation. Its state space is infinite-dimensional, and its basic computational step is autonomous descent on a programmed functional. 

Within this paradigm, ion-irradiated FeRh provides a plausible minimal realization of the local-writing regime. In that realization, spatial control of the local phase bias supplies the encoding channel, AF--FM coexistence supplies the relaxational mechanism, and magnetic-domain imaging supplies the readout channel. The representative tasks above identify a concrete class of computations already expressible in this minimal setting. More general coupling-programmable realizations could broaden the paradigm beyond the local-writing regime and enlarge the class of computations it can implement.

\end{document}